\documentclass[aps,amsmath,showpacs,showkeys,nofootinbib,
superscriptaddress]{revtex4}
\usepackage[dvips]{graphicx}
 
\begin{document} 

\title{\bf Is there a $\pi\Lambda N$ bound state?} 

\author{A. Gal} 
\email{avragal@vms.huji.ac.il} 
\affiliation{Racah Institute of Physics, The Hebrew University,
Jerusalem 91904, Israel}

\author{H. Garcilazo}
\email{humberto@esfm.ipn.mx} 
\affiliation{Escuela Superior de F\' \i sica y Matem\'aticas \\ 
Instituto Polit\'ecnico Nacional, Edificio 9,
07738 M\'exico D.F., Mexico}

\date{\today} 

\begin{abstract} 
We have searched for bound states in the $\pi\Lambda N$ system by 
solving the nonrelativistic Faddeev equations, as well as a relativistic 
version, with input separable $\pi N$, $\pi\Lambda$, and $\Lambda N$ 
interactions. A bound-state solution, driven by the $\Delta(1232)$ 
and $\Sigma(1385)$ $p$-wave meson-baryon resonances, was found in 
the channel $(I,J^P)=(\frac{3}{2},2^+)$, provided the $\Lambda$ laboratory 
momentum at which the $\Lambda N$ $^3S_1$ phase shift becomes negative 
is larger than $p_{\rm lab} \sim 750-800$ MeV/c. Other strange and charmed 
$\pi BB^\prime$ systems that might have bound states of a similar nature 
are listed. 
\end{abstract}



\maketitle 


\section{Introduction} 
\label{sec:int} 

Experimental searches for dibaryons have been inconclusive. 
In the nonstrange sector, pion-initiated reactions and pion-production 
reactions were used to search for low-lying narrow $\pi NN$ resonances 
below the $\Delta N$ threshold, aiming particularly at channels with 
quantum numbers inaccessible to $NN$ configurations \cite{BPB02}. 
Several broad $NN$ resonances are known near the $\Delta N$ 
and $\Delta\Delta$ thresholds and may be attributed to quasibound 
states in these channels, as summarized recently \cite{CBD08}.   
In the strange sector, extensive searches have been conducted 
\cite{Bas97,Col97,Cra97} for the $H$ dibaryon, with strangeness 
$S=-2$ and quantum numbers $(I,J^P)=(0,0^+)$, which originally was 
predicted to lie below the $\Lambda\Lambda$ threshold \cite{Jaf77}. 
Only few dedicated searches for $S=-1$ dibaryons have been reported, 
for low-lying $L=1$ $\Lambda N$ resonances in singlet and triplet 
configurations that were predicted in a quark-model study by Mulders 
{\it et al.} \cite{MAD80} near the $\Sigma N$ threshold, but negative 
results particularly for the singlet resonance were reported in 
$K^-$-initiated experiments \cite{JHK92,CDG92}. 

Here we look for low-lying $S=-1$ dibaryons associated with a `molecular' 
$\pi\Lambda N$ structure, by solving three-body Faddeev equations with 
pairwise phenomenological separable interactions. The $\Lambda N$ system 
is known to be unbound, with $s$-wave forces in both singlet and triplet 
states that are overall attractive and which yield scattering lengths of 
order $-2$ fm \cite{RYa06}. The question is whether or not the pion is able 
to bind an $s$-wave $\Lambda N$ pair within a $\pi\Lambda N$ bound state, 
or a resonance. Since the $s$-wave $\pi N$ and $\pi\Lambda$ forces are very 
weak \cite{Wei66}, we consider the $p$-wave resonances $\Delta(1232)$ 
$(\frac{3}{2},\frac{3}{2}^+)$ and $\Sigma(1385)$ $(1,\frac{3}{2}^+)$, 
respectively, thus studying the $\pi\Lambda N$ three-body system with 
$s$-wave baryons and a $p$-wave pion in a $(\frac{3}{2},2^+)$ state, 
where the $\Lambda N$ subsystem is necessarily in the $^3S_1$ configuration. 
For first orientation we neglect the ${^3S_1}-{^3D_1}$ channel coupling 
which becomes important near and above the $\Sigma N$ threshold. 

For all three partitions of this $(\frac{3}{2},2^+)$ state of the 
$\pi\Lambda N$ system into an interacting pair and a spectator, the orbital 
angular momenta, spins, and isospins couple to their maximal values and, 
therefore, the spin and isospin recoupling coefficients are equal to one. 
This three-body state is likely to represent a state with maximum possible 
attraction. Furthermore, the fact that the spin and isospin recoupling 
coefficients are equal to one allows for a formal reduction of the present 
three-body problem to that of three spinless (and isospinless) particles. 
We comment that a similar choice of $(I,J^P)=(2,2^+)$ for $\pi NN$, with 
each $\pi N$ pair interacting in the $\Delta(1232)$-resonance 
$(\frac{3}{2},\frac{3}{2}^+)$ channel, is impossible since a two-nucleon 
$I=1,{^3S_1}$ state is forbidden by the Pauli principle. 

Since we are interested in the bound-state region of the $\pi\Lambda N$ 
system, it is justified in first approximation to neglect the coupling 
to the higher-mass systems $\bar K N N$, $\pi\Sigma N$ and $K\Xi N$. 
The effect of the coupling to these higher-mass channels will be partly 
taken into account by adjusting the interactions within the 
$\pi\Lambda N$ system to the available experimental information on the 
two-body subsystems. Less justified is the neglect of the coupling to 
the lower-mass $\Sigma N$ system, with a threshold about 60 MeV below that 
of $\pi\Lambda N$. This coupling renders $\pi\Lambda N$ bound states into 
quasibound states through shifting and broadening the zero-width bound 
states obtained when the coupling is disregarded, unless the binding energy 
exceeds approximately 60 MeV and the $\pi\Lambda N$ state is genuinely bound. 
In the present, exploratory calculation we ignore the coupling to $\Sigma N$. 
Potential models generally yield fairly weak $\Sigma N$ interaction in the 
relevant ${^1D_2}$ and ${^3D_2}$ configurations \cite{RYa06}. The quark model 
of Ref.~\cite{MAD80} does not have any $(\frac{3}{2},2^+)$ $S=-1$ dibaryon 
candidate in the vicinity of the $\pi\Lambda N$ threshold and below it. 

The plan of this paper is as follows. In Sec.~\ref{sec:NR} we discuss the 
choice of two-body interactions and the three-body Faddeev equations solved 
in the nonrelativistic case, and report on the binding energies calculated 
for the $(\frac{3}{2},2^+)$ $\pi\Lambda N$ system. The corresponding analysis 
of, and the binding energies calculated in a relativistic version of the 
three-body model are discussed in Sec.~\ref{sec:R}. The paper ends with 
a brief summary and discussion in Sec.~\ref{sec:sum}, where additional 
strange and charmed $\pi BB^\prime$ systems that might admit bound states 
of a similar nature are listed. 

\section{A nonrelativistic model} 
\label{sec:NR} 

\subsection{The two-body subsystems}
\label{sec:2NR} 

Since both $\pi\Lambda$ and $\pi N$ subsystems are dominated by 
$p$-wave resonances, we assumed a rank-one separable meson-baryon 
interaction 
\begin{equation} 
V_i(p_i,p_i^\prime)=-g_i(p_i)g_i(p_i^\prime)~. 
\label{eq1} 
\end{equation} 
The corresponding two-body $t$-matrix is given by 
\begin{equation}
t_i(p_i,p_i^\prime;E)=-g_i(p_i)\tau_i(E)g_i(p_i^\prime)~, 
\label{eq2} 
\end{equation} 
where $E$ is the energy in the two-body center-of-mass (c.m.) system and 
\begin{equation} 
\tau_i^{-1}(E)= 1+\int_0^\infty p_i^2dp_i \frac{g_i^2(p_i)}
{E-p_i^2/2\eta_i+{\rm i}\epsilon}~, 
\label{eq3} 
\end{equation} 
with $\eta_i=m_jm_k/(m_j+m_k)$, where $\epsilon_{ijk} \neq 0$. 
The form factors $g_i(p_i)$ are chosen of the form 
\begin{equation} 
g_i(p_i)=\sqrt{\gamma_i}p_i(1+p_i^2)e^{-p_i^2/\alpha_i^2}~, 
\label{eq4} 
\end{equation} 
where the two parameters $\gamma_i$ and $\alpha_i$ were adjusted 
to the position and width of the corresponding resonances, as given 
by the Particle Data Group \cite{PDG06}. These parameters are listed 
in Table I for the $\pi N$ and $\pi\Lambda$ subsystems. We also 
constructed a second model of the $\pi N$ interaction of the form 
\begin{equation} 
g_i(p_i)=\sqrt{\gamma_i}p_i[1+(p_i/4.5)^2+(p_i/1.35)^4]e^{-p_i^2/\alpha_i^2}~, 
\label{eq5} 
\end{equation} 
which reproduces in addition the $\pi N$ $P_{33}$ scattering volume. The 
parameters of this model are also given in Table~\ref{tab:piNR}. Note that 
$p_i$ in Eqs.~(\ref{eq4}) and (\ref{eq5}) assumes values in fm$^{-1}$ units.  

\begin{table} 
\caption{Parameters of the pion-baryon separable potentials 
Eqs.~(\ref{eq4}) and (\ref{eq5}), $\alpha_i$ (in fm$^{-1}$) and $\gamma_i$ 
(in fm$^4$), for the nonrelativistic model.} 
\label{tab:piNR} 
\begin{ruledtabular} 
\begin{tabular}{cccccc} 
 & $\alpha_{\pi N}$ & $\gamma_{\pi N}$ & $\alpha_{\pi\Lambda}$ & 
$\gamma_{\pi\Lambda}$ &  \\  \hline 
Eq.~(\ref{eq4}) &  2.021352   &  0.02116  &   2.523999  &   0.00564  & \\ 
Eq.~(\ref{eq5}) &  1.560768  &  0.06244  &    $ - $    &   $ - $    & \\ 
\end{tabular} 
\end{ruledtabular} 
\end{table}

For the $^3S_1$ $\Lambda N$ subsystem we assume a rank-two separable 
potential consisting of both attractive and repulsive terms 
\begin{equation}
V_i(p_i,p_i^\prime)=-g_i^a(p_i)g_i^a(p_i^\prime)+g_i^r(p_i)g_i^r(p_i^\prime)~. 
\label{eq6} 
\end{equation} 
The corresponding two-body $t$-matrix is given by 
\begin{equation} 
t_i(p_i,p_i^\prime;E)=-\sum_{\alpha=a,r}\sum_{\beta=a,r} 
g_i^\alpha(p_i)\tau_i^{\alpha\beta}(E)g_i^\beta(p_i^\prime)~, 
\label{eq7} 
\end{equation} 
where 
\begin{equation} 
\tau_i^{ar}(E)=\tau_i^{ra}(E)=\frac{G_i^{ar}(E)}{[1+G_i^{aa}(E)] 
[1-G_i^{rr}(E)]+[G_i^{ar}(E)]^2}~, 
\label{eq8} 
\end{equation} 
\begin{equation} 
\tau_i^{aa}(E)=\frac{1-G_i^{rr}(E)}{[1+G_i^{aa}(E)]
[1-G_i^{rr}(E)]+[G_i^{ar}(E)]^2}~, 
\label{eq9} 
\end{equation} 
\begin{equation} 
\tau_i^{rr}(E)=-\frac{1+G_i^{aa}(E)}{[1+G_i^{aa}(E)] 
[1-G_i^{rr}(E)]+[G_i^{ar}(E)]^2}~, 
\label{eq10} 
\end{equation} 
\begin{equation} 
G_i^{\alpha\beta}(E)=\int_0^\infty p_i^2dp_i \frac{g_i^\alpha(p_i) 
g_i^\beta(p_i)}{E-p_i^2/2\eta_i+{\rm i}\epsilon}~.
\label{eq11} 
\end{equation} 
The form factors $g_i^\beta(p_i)$ 
are chosen to be of the Yamaguchi form 
\begin{equation} 
g_i^\beta(p_i)=\frac{\sqrt{\gamma_\beta}}{p_i^2+\alpha_\beta^2}~~~~ 
(\beta=a,r)~,
\label{eq12} 
\end{equation} 
where the parameters $\alpha_a$, $\gamma_a$, $\alpha_r$, and $\gamma_r$ are 
adjusted to reproduce given values of the $\Lambda N$ $^3S_1$ scattering 
length and effective range for different values of the $\Lambda$ laboratory 
momentum $p_{\rm lab}^{(0)}$ at which the $^3S_1$ $\Lambda N$ phase shift 
becomes negative, changing sign from attraction at low momentum to repulsion 
at high momentum (as discussed in Sec.~\ref{sec:NResults}). The values of 
the scattering length and effective range adopted here are $a=-1.86$~fm 
and $r_0=3.13$~fm, respectively, corresponding to model ESC04d of 
Ref.~\cite{RYa06}. These values are very close to those in models 
NSC97e,f \cite{RSY99} which have been widely used in $\Lambda$-hypernuclear 
calculations.

\subsection{The three-body system} 
\label{sec:3NR} 

Since all the angular momenta, spins, and isospins are coupled 
to their maximal values, the recoupling coefficients of spin and 
isospin are equal to one, and the Faddeev equations depend only on 
the orbital angular momenta $\vec\ell,~\vec\lambda,~\vec L$, where 
$\vec L=\vec\ell+\vec\lambda$, with $L=1$. The values of $\vec\ell$ 
and $\vec\lambda$ are $\ell=1$, $\lambda=0$ for configurations in 
which the pion interacts with one of the baryons while the other 
baryon is a spectator, and $\ell=0$, $\lambda=1$ for the configuration 
in which the two baryons interact while the pion is a spectator. 

Below we denote the $\Lambda$ hyperon as particle 1, the nucleon as 
particle 2, and the pion as particle 3. Thus, the Faddeev equations for 
the bound-state problem, using the separable potentials (\ref{eq1}) and 
(\ref{eq6}), are 
\begin{equation} 
T_i(q_i)=-\tau_i(E-q_i^2/2\nu_i)\sum_{j=1}^2\int_0^\infty dq_j^\prime 
H_{ij}(q_i,q_j^\prime)T_j(q_j^\prime)~~~~~(i=1,2)~, 
\label{eq13} 
\end{equation} 
with $\nu_i=m_i(m_j+m_k)/(m_i+m_j+m_k)$, where $\epsilon_{ijk} \neq 0$, 
and 
\begin{equation} 
H_{ij}(q_i,q_j^\prime)=(1-\delta_{ij})K_{ij}(q_i,q_j^\prime)-
\sum_{\alpha=a,r}\sum_{\beta=a,r}\int_0^\infty dq_3 
K_{i3}^\alpha(q_i,q_3)\tau_3^{\alpha\beta}(E-q_3^2/2\nu_3) 
K_{3j}^\beta(q_3,q_j^\prime)~.
\label{eq14} 
\end{equation} 
The kernels in Eq.~(\ref{eq14}) are given by 
\begin{equation} 
K_{12}(q_1,q_2)=\frac{1}{2}q_1q_2\int_{-1}^1 d{\cos}\theta\,
\frac{g_1(p_1)(\hat p_1\cdot\hat p_2)g_2(p_2)}
{E-p_2^2/2\eta_2-q_2^2/2\nu_2}~, 
\label{eq15} 
\end{equation} 
\begin{equation} 
K_{31}^\alpha(q_3,q_1)=\frac{1}{2}q_1q_3 
\int_{-1}^1 d{\cos}\theta\, 
\frac{g_3^\alpha(p_3)(\hat q_3\cdot\hat p_1)g_1(p_1)}
{E-p_1^2/2\eta_1-q_1^2/2\nu_1}~,
\label{eq16} 
\end{equation} 
\begin{equation} 
K_{23}^\alpha(q_2,q_3)=\frac{1}{2}q_2q_3 
\int_{-1}^1 d{\cos}\theta\, 
\frac{g_2(p_2)(\hat p_2\cdot\hat q_3)g_3^\alpha(p_3)}
{E-p_3^2/2\eta_3-q_3^2/2\nu_3}~. 
\label{eq17} 
\end{equation} 
From the three previous expressions one obtains the other three 
that correspond to $K_{ji}(q_j,q_i)=K_{ij}(q_i,q_j)$. 
One can calculate $p_i$, $p_j$, ($\hat p_1\cdot\hat p_2$), 
($\hat q_3\cdot\hat p_1$), and ($\hat p_2\cdot\hat q_3$) by using 
\begin{equation} 
\vec p_i=-\vec q_j-a_{ij}\vec q_i~,~~~~\vec p_j=\vec q_i+a_{ji}\vec q_j~,
\label{eq18} 
\end{equation} 
where $(i,j)$ is a cyclic pair, ${\cos}\theta=\hat q_i\cdot\hat q_j$, and 
\begin{equation} 
a_{ij}=\frac{\eta_i}{m_k}~,~~~~a_{ji}=\frac{\eta_j}{m_k}~.  
\label{eq19}
\end{equation} 

In order to find the bound-state solutions of Eq.~(\ref{eq13}), integrals 
were replaced by sums applying numerical integration quadrature. In this way 
Eq.~(\ref{eq13}) becomes a set of homogeneous linear equations. This set has 
solutions only if the determinant of the matrix of its coefficients (the 
Fredholm determinant) vanishes at certain energies. Thus, the procedure to 
find the bound-state energies of the three-body system simply consists of 
searching for the zeros of the Fredholm determinant on the real energy axis. 
Some limiting situations are discussed in the Appendix. 

\subsection{Results} 
\label{sec:NResults} 

\begin{table} 
\caption{Parameters of the $\Lambda N$ $^3S_1$ potentials (\ref{eq12}) 
$\alpha_\beta$ (in fm$^{-1}$), $\gamma_\beta$ (in fm$^{-2}$) in the 
nonrelativistic model for $a=-1.86$ fm, $r_0=3.13$ fm, and the binding 
energies $B_{\pi\Lambda N}$ (in MeV) of the three-body $\pi\Lambda N$ 
system calculated using the $\pi N$ and $\pi \Lambda$ potential parameters 
listed in Table~\ref{tab:piNR}, Eq.~(\ref{eq4}) [the $B_{\pi\Lambda N}$ 
values in parentheses correspond to the $\pi N$ parameters listed in 
Table~\ref{tab:piNR}, Eq.~(\ref{eq5})]. The momentum $p_{\rm lab}^{(0)}$ 
(in MeV/c) is the laboratory $\Lambda$ momentum at which the $\Lambda N$ 
$^3S_1$ phase shift becomes negative.} 
\label{tab:NResults} 
\begin{ruledtabular} 
\begin{tabular}{cccccccc}
& $\alpha_a$ & $\gamma_a$ &  $\alpha_r$  & $\gamma_r$  
& $p_{\rm lab}^{(0)}$  & $B_{\pi\Lambda N}$ & \\ 
\hline 
     &   1.437 &  0.4179  & $ - $  & $ - $  & $ - $ & 140 &  \\
     &   1.6   &  0.8118  &   4.0  &  5.54  &  1184 & 111 &  \\
     &   1.6   &  0.8053  &   6.0  &  26.0  &  1069 & 96 &  \\
     &   1.6   &  0.8064  &   8.0  &  86.0  &  1045 & 86 &  \\
     &   1.7   &  1.195   &   4.0  &  10.0  &  975  & 92 &  \\
     &   1.7   &  1.186   &   6.0  &  51.0  &  910  & 66 &  \\
     &   1.7   &  1.190   &   8.0  & 190.0  &  899  & 52 &  \\
     &   1.8   &  1.735   &   4.0  &  15.5  &  877  & 72 (67) & \\
     &   1.8   &  1.718   &   6.0  &  86.0  &  834  & 38 (37) & \\
     &   1.8   &  1.745   &   8.0  &  405.0 &  826  & 21 (23) & \\
     &   1.9   &  2.513   &   4.0  &  22.7  &  814  & 51 &  \\
     &   1.9   &  2.501   &   6.0  & 145.0  &  784  &  9 &  \\
     &   1.9   &  2.573   &   8.0  & 1150.0 &  779  & unbound & \\
     &   2.0   &  3.588   &   4.0  &   31.4 &  777  & 31 &  \\
     &   2.0   &  3.602   &   6.0  &  244.0 &  753  & unbound & \\
     &   2.1   &  5.125   &   4.0  &   42.9 &  748  & 10 &  \\
     &   2.2   &  7.311   &   4.0  &   58.0 &  728  & unbound & \\ 
\end{tabular} 
\end{ruledtabular} 
\end{table} 

In the last column of Table~\ref{tab:NResults}, we list the calculated 
binding energies $B_{\pi\Lambda N}$ of the $\pi\Lambda N$ system in the 
$(I,J^P)=(\frac{3}{2},2^+)$ channel, for the $\pi\Lambda$ and $\pi N$ 
interactions recorded in Table~\ref{tab:piNR} and the various models of the 
$\Lambda N$ interaction also listed in Table~\ref{tab:NResults}. Most of the 
results are given for the choice Eq.~(\ref{eq4}) of the $\pi N$ form factor, 
except for the $\alpha_a=1.8$~fm$^{-1}$ runs for which listed in parentheses 
are also the binding energies obtained using the other choice Eq.~(\ref{eq5}). 
The dependence on the type of $\pi N$ form factor is seen to be rather weak. 
We also checked the sensitivity to the strength parameter 
$\gamma_{\pi\Lambda}$; for example, the $\pi\Lambda N$ bound state for the 
case $B_{\pi\Lambda N}=51$ MeV listed in the table disappears as soon as the 
standard value $\gamma_{\pi\Lambda}=0.00564$~fm$^4$ from Table~\ref{tab:piNR} 
is decreased to 0.00524~fm$^4$. The dependence on the $\Lambda N$ interaction 
is shown in detail in Table~\ref{tab:NResults}. Essentially, the various 
$\Lambda N$ models differ from each other by the amount of repulsion they 
contain. For a given value of range parameter $\alpha_a^{-1}$ for the 
attractive $\Lambda N$ component, the calculated binding energy decreases as 
the repulsive component gets pushed inside and requires a larger strength. 
For a given value of range parameter $\alpha_r^{-1}$ for the repulsive 
component, the calculated binding energy decreases as the attractive 
component gets pushed inside, or equivalently as one lowers the momentum 
where the $\Lambda N$ $^3S_1$ phase shift changes sign from positive 
(attraction) to negative (repulsion) values. It is seen that the bound 
state persists as long as this $\Lambda$ laboratory momentum 
$p_{\rm lab}^{(0)}$ is larger than about $750-800$ MeV/c. Incidentally, 
this is precisely the range of momenta at which the $\Lambda N$ $^3S_1$ 
phase shift goes through zero in Nijmegen $YN$ potential models that 
relegate the ${^3S_1}-{^3D_1}$ attraction near and above the $\Sigma N$ 
threshold to the $^3D_1$ channel \cite{RSY99}.

\section{A relativistic model} 
\label{sec:R} 

Since the binding energies calculated nonrelativistically, for some of 
the cases listed in Table~\ref{tab:NResults} are a sizable fraction 
of the pion mass, it appears necessary to take into account relativistic 
effects. Therefore, we will reformulate our model in terms of 
a relativistic on-mass-shell-spectator formalism \cite{Gro82,Gar87,SGF97}. 
In this formalism one starts with the Bethe-Salpeter equation for three 
particles which is set in a Faddeev form. The four-vector equations are 
then reduced to three-vector equations similar to the nonrelativistic 
Faddeev equations by putting all the spectator particles on the mass 
shell \cite{Gar87}. 

In order to reach a relativistic generalization of Eq.~(\ref{eq13}) 
we make two approximations. First, the negative-energy components of the 
fermion propagators are neglected; and second, the spin degrees of freedom 
are treated nonrelativistically by means of Racah coefficients (which are 
equal to one, as pointed out above). 
These two approximations are reasonable since the two fermions $\Lambda$ 
and $N$ are very heavy compared with the pion. Thus, as pointed out in 
the Introduction, our model formally  reduces to that of three 
spinless (and isospinless) particles interacting by pairwise separable 
interactions. 

\subsection{The two-body subsystems} 
\label{sec:2R} 

In order to fit the $p$-wave resonance energy and width in the $\pi\Lambda$ 
and $\pi N$ subsystems we considered the two-body Bethe-Salpeter equation for 
the pair $jk$ with particle $j$ (here the pion) on the mass shell interacting 
through a rank-one separable interaction defined by Eqs.~(\ref{eq1}) and 
(\ref{eq4}). Recall that $p_i$, the magnitude of the relative three-momentum 
of the pair in the c.m. system, is Lorentz invariant since it is expressible 
in terms of the invariant mass of the relative momentum four-vector. 
The corresponding two-body $t$-matrix in the c.m. system is given by 
\begin{equation} 
t_i(p_i,p_i^\prime;\omega_0)=-g_i(p_i)\tau_i(\omega_0)g_i(p_i^\prime)~, 
\label{eq20} 
\end{equation} 
where $\omega_0$ is the invariant mass of the two-body subsystem and 
\begin{equation} 
\tau_i^{-1}(\omega_0)= 1+\int_0^\infty \frac{p_i^2dp_i}{2\omega_j} 
\frac{g_i^2(p_i)}{(\omega_0-\omega_j)^2-\omega_k^2+{\rm i}\epsilon}~, 
\label{eq21} 
\end{equation} 
with $\omega_j=\sqrt{m_j^2+p_i^2}$ and $\omega_k=\sqrt{m_k^2+p_i^2}$. The 
parameters of these separable potentials are given in Table~\ref{tab:piR}. 
We did not pursue the option of keeping the respective baryon on mass shell, 
with an off-shell pion, because of the appearance of a persistent unphysical 
two-body bound state for this choice. 

\begin{table} 
\caption{Parameters of the pion-baryon separable potential 
Eq.~(\ref{eq4}), $\alpha_i$ (in fm$^{-1}$) and $\gamma_i$ (in fm$^2$), 
for the relativistic model with on-mass-shell $\pi$ meson.} 
\label{tab:piR} 
\begin{ruledtabular} 
\begin{tabular}{cccccc} 
 & $\alpha_{\pi N}$ & $\gamma_{\pi N}$ & $\alpha_{\pi\Lambda}$ & 
$\gamma_{\pi\Lambda}$  &  \\  \hline
Eq.~(\ref{eq4}) & 2.231357 & 0.219260 & 2.720821 & 0.083916 & \\
\end{tabular} 
\end{ruledtabular} 
\end{table} 

For the $\Lambda N$ subsystem we again used a rank-two separable 
potential defined by Eqs.~(\ref{eq6}) and (\ref{eq12}) so that the 
two-body $t$-matrix is given by Eqs.~(\ref{eq7})-(\ref{eq10}) with 
$E$ replaced by $\omega_0$ and $G_i^{\alpha\beta}(E)$ of Eq.~(\ref{eq11}) 
replaced by 
\begin{equation} 
G_i^{\alpha\beta}(\omega_0)= \int_0^\infty \frac{p_i^2dp_i}{2\omega_j}
\frac{g_i^\alpha(p_i)g_i^\beta(p_i)}
{(\omega_0-\omega_j)^2-\omega_k^2+{\rm i}\epsilon}~. 
\label{eq22} 
\end{equation} 
The parameters of these separable potentials are listed below in 
Sec.~\ref{sec:Results}.

\subsection{The three-body system} 
\label{sec:3R} 

The integral equations for the three-body problem are given by 
\begin{equation}
T_i(q_i)=-\tau_i(W_0;q_i)\sum_{j=1}^2\int_0^{q^{(j)}_{\rm max}} 
dq_j^\prime H_{ij}(q_i,q_j^\prime)T_j(q_j^\prime)~~~~~(i=1,2)~, 
\label{eq23} 
\end{equation} 
where $W_0$ is the invariant mass of the three-body system. 
The upper limit of integration, 
\begin{equation} 
q^{(j)}_{\rm max}=\frac{W_0^2-m_j^2}{2W_0}~, 
\label{eq24}
\end{equation} 
is the momentum at which the invariant mass of the two-body subsystem 
$j$ is equal to zero so that it then recoils with the speed of light 
\cite{SGF97}. The entity $\tau_i(W_0;q_i)$ corresponds to the $t$-matrix 
(\ref{eq20})-(\ref{eq21}) in an arbitrary frame where the spectator 
particle $i$ (which is on-mass-shell) has momentum $\vec q_i$, particle 
$j$ (which has also been put on-mass-shell) has momentum $\vec q_j$ 
and particle $k$ (which is off the mass shell) has momentum 
$-\vec q_i-\vec q_j$. It is given by 
\begin{equation} 
\tau_i^{-1}(W_0;q_i)= 1+\frac{1}{2}\int_{-1}^1d{\cos}\theta
\int_0^\infty\frac{q_j^2dq_j}{2\omega_j}\frac{g_i^2(p_i)}
{(W_0-\omega_i-\omega_j)^2-\omega_k^2+{\rm i}\epsilon}~,
\label{eq25}
\end{equation} 
with 
\begin{equation} 
\omega_i=\sqrt{m_i^2+q_i^2}~,~~~~\omega_j=\sqrt{m_j^2+q_j^2}~,
\label{eq26} 
\end{equation} 
\begin{equation} 
\omega_k=\sqrt{m_k^2+q_i^2+q_j^2+2q_iq_j{\cos}\theta}~. 
\label{eq27} 
\end{equation} 
The magnitude of the relative three-momentum $\vec p_i$ is 
a Lorentz invariant given by 
\begin{equation} 
p_i^2=\frac{(P_{jk}^2+m_j^2-k_k^2)^2}{4P_{jk}^2}-m_j^2~, 
\label{eq28}
\end{equation} 
where $P_{jk}=k_j+k_k$ is the total four-momentum of the pair $jk$ 
and $k_k$ is the four-momentum of particle $k$, {\it i.e.}, 
\begin{equation} 
P_{jk}^2=(W_0-\omega_i)^2-q_i^2~,
\label{eq29} 
\end{equation} 
\begin{equation} 
k_k^2=(W_0-\omega_i-\omega_j)^2
-q_i^2-q_j^2-2q_iq_j{\cos}\theta~.
\label{eq30} 
\end{equation} 
Eq.~(\ref{eq25}) reduces to Eq.~(\ref{eq21}) when $q_i=0$. 
Similar expressions apply to the relativistic version of the 
$\Lambda N$ $t$-matrix in an arbitrary frame 
$\tau_3^{\alpha\beta}(W_0;q_3)$. 

The kernel of Eq.~(\ref{eq23}) is given by Eqs.~(\ref{eq14})-(\ref{eq17}), 
where the upper limit $\infty$ in the integral of Eq.~(\ref{eq14}) is 
replaced by $q_{\rm max}^{(3)}$, and the following substitutions are made:   
\begin{equation} 
\frac{1}{E-p_j^2/2\eta_j-q_j^2/2\nu_j} ~\to~ 
\frac{1}{2\omega_j}\frac{1}{(W_0-\omega_i-\omega_j)^2-\omega_k^2}~,
\label{eq31} 
\end{equation}
\begin{equation} 
a_{ij}~ \to~ \frac{W_i^2-q_i^2+m_j^2-k_k^2+2\omega_j\sqrt{W_i^2-q_i^2}}
{2\sqrt{W_i^2-q_i^2}~(W_i+\sqrt{W_i^2-q_i^2}~)}~,
\label{eq32} 
\end{equation}
\begin{equation} 
a_{ji}~ \to~ \frac{W_j^2-q_j^2+m_i^2-k_k^2+2\omega_i\sqrt{W_j^2-q_j^2}}
{2\sqrt{W_j^2-q_j^2}~(W_j+\sqrt{W_j^2-q_j^2}~)}~,
\label{eq33} 
\end{equation}
\begin{equation} 
W_i=W_0-\omega_i~,~~~~W_j=W_0-\omega_j~. 
\label{eq34} 
\end{equation} 
Eq.~(\ref{eq31}) is the propagator when the spectator particles $i$ 
and $j$ are on-mass-shell and the exchanged particle $k$ is 
off-mass-shell. Eqs.~(\ref{eq32})-(\ref{eq34}) correspond to the 
relativitistic kinematics with particle $k$ off the mass shell.

\subsection{Results} 
\label{sec:Results} 

\begin{table} 
\caption{Parameters of the $\Lambda N$ $^3S_1$ potentials (\ref{eq12}) 
$\alpha_\beta$ (in fm$^{-1}$), $\gamma_\beta$ (in fm$^{-4}$) in the 
relativistic model with on-mass-shell nucleon, for $a=-1.86$ fm, 
$r_0=3.13$ fm, and the binding energies $B_{\pi\Lambda N}$ (in MeV) 
of the three-body $\pi\Lambda N$ system calculated using the $\pi N$ 
and $\pi \Lambda$ potential parameters listed in Table~\ref{tab:piR}. 
The momentum $p_{\rm lab}^{(0)}$ (in MeV/c) is the laboratory $\Lambda$ 
momentum at which the $\Lambda N$ $^3S_1$ phase shift becomes negative.} 
\label{tab:Results} 
\begin{ruledtabular} 
\begin{tabular}{cccccccc} 
& $\alpha_a$ & $\gamma_a$ &  $\alpha_r$  & $\gamma_r$ 
& $p_{\rm lab}^{(0)}$  & $B_{\pi\Lambda N}$ & \\ 
\hline
   &  2.0     &  318.2  &  4.0  &    2270  &  866  & 152 &  \\
   &  2.0     &  309.2  &  6.0  &   12100  &  823  & 93 &  \\
   &  2.0     &  313.0  &  8.0  &   54500  &  813  & 69 &  \\
   &  2.1     &  446.9  &  4.0  &    3080  &  823  & 121 &  \\
   &  2.1     &  434.3  &  6.0  &   18000  &  788  & 59 &  \\
   &  2.1     &  440.8  &  8.0  &  105000  &  783  & 35 &  \\
   &  2.2     &  626.6  &  4.0  &    4100  &  791  & 94 &  \\
   &  2.2     &  599.1  &  6.0  &   25800  &  768  & 31 &  \\
   &  2.2     &  632.5  &  8.0  &  350000  &  756  & unbound &  \\
   &  2.3     &  878.5  &  4.0  &    5400  &  766  & 69 &  \\
   &  2.3     &  845.8  &  6.0  &   40700  &  746  & unbound &  \\
   &  2.4     & 1217    &  4.0  &    6930  &  750  & 48 &  \\
   &  2.4     & 1189    &  6.0  &   68000  &  733  & unbound &  \\
   &  2.5     & 1728    &  4.0  &    9200  &  730  & 21 &  \\    
   &  2.6     & 2354    &  4.0  &   11400  &  728  & 6 &  \\
\end{tabular}
\end{ruledtabular} 
\end{table}

In the last column of Table~\ref{tab:Results}, we list the calculated 
binding energies $B_{\pi\Lambda N}$ of the $\pi\Lambda N$ system in the 
$(I,J^P)=(\frac{3}{2},2^+)$ channel, for the $\pi\Lambda$ and $\pi N$ 
interactions recorded in Table~\ref{tab:piR} and the various models of 
the $\Lambda N$ interaction listed also in Table~\ref{tab:Results}. 
The dependence of the calculated binding energies on the ranges of the 
repulsive and attractive components of the $\Lambda N$ interaction is 
similar to that found in the nonrelativistic calculations. 
A bound state in the relativistic model persists as long as the $\Lambda$ 
laboratory momentum at which the $\Lambda N$ phase shift becomes negative, 
$p_{\rm lab}^{(0)}$, is larger than about 750 MeV/c. 
A comparison between Tables~\ref{tab:NResults} and \ref{tab:Results} 
reveals that the relativistic model provides more attraction than the 
nonrelativistic one, in agreement with the slower increase of kinetic 
energy with momentum when relativistic kinematics is applied.

\section{Summary and Discussion} 
\label{sec:sum}

We have used a nonrelativistic separable potential model and 
a relativistic version of it, solving three-body Faddeev equations, 
to search for $\pi\Lambda N$ bound states. In both models we found 
that a $(I,J^P)=(\frac{3}{2},2^+)$ bound state is likely to exist, 
provided the $\Lambda$ laboratory momentum $p_{\rm lab}^{(0)}$ 
at which the ${^3S_1}$ $\Lambda N$ phase shift becomes negative 
is larger than about $750-800$ MeV/c. This agrees with the range 
of momenta at which Nijmegen $YN$ potential models, where applicable 
\cite{RSY99}, predict that the ${^3S_1}$ $\Lambda N$ phase shift goes 
through zero. The J\"{u}lich '04 model \cite{HMe05} and the recent 
chiral EFT approach \cite{PHM06} predict that $p_{\rm lab}^{(0)}>900$ 
MeV/c, so that the existence of a $\pi\Lambda N$ bound state in 
these models appears robust. The Nijmegen and J\"{u}lich $YN$ 
potential models differ considerably from each other within 
the $\Lambda N$ $J^P=1^+$ coupled channels also in the behavior 
of the ${^3D_1}$ phase shift. The ${^3S_1}-{^3D_1}$ coupling was 
neglected in the present exploratory three-body calculation, 
a neglect that might be justified in applications of the 
J\"{u}lich models where both the coupling and the size of the 
${^3D_1}$ phase shift that builds up above the $\Sigma N$ 
threshold at $p_{\rm lab} \approx 630$ MeV/c are weaker than 
in the Nijmegen models. However, all these $YN$ models have been 
constructed to fit primarily low-energy scattering data which do not 
unambiguously constrain the short-range behavior of the ${^3S_1}$ 
$\Lambda N$ system. The extent to which the two-body short-range 
repulsion varies between `soft' to `hard' is crucial for the three-body 
system's ability to bind, with the $p$-wave pion maximizing its 
attraction to each one of the baryons simultaneously.    

More realistic three-body calculations 
will have to include $\Sigma$ hyperons, extending the $\Lambda N$ 
channel into ${^3S_1}-{^3D_1}$ $\Lambda N-\Sigma N$ coupled channels, 
and the $\pi\Lambda$ channel into $\pi\Lambda-\pi\Sigma$ coupled 
channels. Although the $I=1$ $\bar K N$ channel also couples to these 
$\pi Y$ coupled channels, in first approximation the three-body 
$\bar K NN$ channel is decoupled from the $\pi YN$ coupled channels for 
$(I,J^P)=(\frac{3}{2},2^+)$ owing to the restrictions imposed by the 
Pauli principle on the two nucleons. 

To search experimentally for a possible $I=\frac{3}{2},~J^P=2^+$ 
$\pi\Lambda N$ dibaryon bound state or resonance, which we denote 
by $\cal D$, one could try in-flight $(K^-,\pi^+)$ or $(\pi^-,K^+)$ 
reactions on a deuteron target: 
\begin{equation} 
K^- + d \to  {\cal D}^- + \pi^+~, 
\label{eq35}
\end{equation} 
\begin{equation} 
\pi^- + d \to {\cal D}^- + K^+~. 
\label{eq36} 
\end{equation} 
These reactions lead automatically to the required value of isospin 
$I=\frac{3}{2}$ for the $\cal D$ dibaryon. The values required for 
spin-parity, $J^P=2^+$, are also allowed. In terms of a coupled 
$\Sigma^- n$ system, the orbital angular momentum and Pauli-spin 
are approximately conserved, resulting in two possibilities: 
${^3D_2}$ and ${^1D_2}$. These could be explored by choosing 
an incident momentum and a meson scattering angle where the 
$K^- + p \to \Sigma^- + \pi^+$ or $\pi^- + p \to \Sigma^- + K^+$ 
underlying reactions are largely non-spin-flip ($\to {^3D_2}$) or 
have a nonnegligible spin-flip component ($\to {^1D_2}$). 
These experiments would be feasible at J-PARC.  

The three-body calculations reported here for the $S=-1$ $\pi \Lambda N$ 
system may be extended to other three-body systems of the type $\pi B_1 B_2$, 
with $J^P=2^+$ and a maximum value of isospin, consisting of a $p$-wave pion 
and $\frac{1}{2}^+$ baryons in a relative $s$-wave state. This precludes 
identical baryons: $B_1 \neq B_2$. Candidates may be classified as follows: 

\begin{itemize} 
\item $S=-2,-3$ strange systems obtained by substituting the SU(3)-octet $\Xi$ 
hyperon for the $\Lambda$ hyperon or for the nucleon in the $\pi\Lambda N$ 
three-body system, leading to $\pi\Xi N$ and $\pi\Lambda\Xi$, respectively. 
The new $\pi \Xi$ $p$-wave resonance here is the $\frac{3}{2}^+$ $\Xi(1530)$ 
belonging to the same SU(3) decuplet which contains the $\Delta(1232)$ and 
the $\Sigma(1385)$ considered in the present work. 
\item $C=+1$ charmed systems made out of a pion, SU(3)-octet baryon 
(excluding the $\Sigma$ hyperon) and $\frac{1}{2}^+$ charmed baryon 
(of the lowest mass for a given strangeness): 
\begin{equation} 
{\pi}N\Lambda_c(2286)~,~~~~{\pi}N\Xi_c(2470)~,~~~~{\pi}N\Omega_c(2700)~,
\end{equation} 
\begin{equation} 
\pi\Lambda\Lambda_c(2286)~,~~~~\pi\Lambda\Xi_c(2470)~,~~~~
\pi\Lambda\Omega_c(2700)~, 
\end{equation} 
\begin{equation} 
\pi\Xi\Lambda_c(2286)~,~~~~\pi\Xi\Xi_c(2470)~,~~~~\pi\Xi\Omega_c(2700)~. 
\end{equation} 
\item $C=+2$ charmed systems made out of a pion and two $\frac{1}{2}^+$ 
singly charmed baryons, each of the lowest mass for a given strangeness:
\begin{equation} 
\pi\Lambda_c(2286)\Xi_c(2470)~,~~~~\pi\Lambda_c(2286)\Omega_c(2700)~,~~~~
\pi\Xi_c(2470)\Omega_c(2700)~. 
\end{equation} 
\end{itemize} 
Note the appearance of the $\frac{1}{2}^+$ $\Omega_c$ baryon, of quark 
structure $ssc$. In the case of charmed baryons, the $p$-wave non-charmed 
SU(3)-decuplet $\frac{3}{2}^+$ resonances are replaced by charmed 
SU(3)-sextet members of the same extended SU(4) {\bf 20}-plet: 
\begin{equation} 
\Sigma(1385)\to\Sigma_c(2520)~,~~~~\Xi(1530)\to\Xi_c(2645)~,~~~~
\Omega(1670)\to\Omega_c(2770)~. 
\end{equation} 
Here we limited listing to singly-charmed baryons. The only observation 
we wish to make on a future charmed bound-state study is that the 
$\pi N \Lambda_c(2286)$ threshold lies {\it below} $N \Sigma_c(2455)$, 
where $\Sigma_c(2455)$ is the lowest lying known $\Sigma_c$, with assumed 
$J^P=\frac{1}{2}^+$. Therefore, if $\pi N \Lambda_c(2286)$ is bound, it will 
decay only by weak interactions. Hopefully, the study of these, and 
other charmed dibaryons will become feasible in due course. 

\section*{Acknowledgments} 
We thank Marek Karliner for bringing to our attention possible implications 
of the present calculation to binding $\pi N \Lambda_c$. 
This work was supported in part by the Israel Science Foundation grant 
757/05 and by COFAA-IPN (M\'exico).

\section*{Appendix: Limiting Faddeev solutions for $\pi\Lambda N$, $\pi NN$ 
and $\pi\Lambda\Lambda$} 

It is interesting to solve the coupled Faddeev Eqs.~(\ref{eq13}) in the 
limit of vanishing baryon-baryon interaction, $\tau_3^{\alpha\beta}=0$. 
Eq.~(\ref{eq14}) reduces then to $H_{ij}=(1-\delta_{ij})K_{ij}$, 
for $i,j=1,2$, so that Eqs.~(\ref{eq13}) become 
\begin{equation} 
T_i = -\tau_i K_{ij} \ast T_j~,~~~~(i \neq j)~, 
\end{equation} 
where the asterisk stands for convolution. Bound states are obtained by 
searching for zeros of the Fredholm determinant corresponding to the 
operator $(1-\tau_1 K_{12} \tau_2 K_{21})$. Using $\pi N$ and $\pi\Lambda$ 
interaction parameters from Table~\ref{tab:piNR}, Eq.~(\ref{eq4}), 
a robust bound state is found at $B_{\pi\Lambda N}=110$~MeV. 
From Table~\ref{tab:NResults} we learn that a fully attractive 
$\Lambda N$ interaction leads to a higher value of $B_{\pi\Lambda N}$, 
and that the introduction of a repulsive component quickly lowers the 
calculated $B_{\pi\Lambda N}$ values below that for a noninteracting 
$\Lambda N$ pair.   

Next, let's make the two baryons identical as far as their mass, 
spin-parity $\frac{1}{2}^+$, and interaction with the pion are concerned. 
Then, $\tau_1=\tau_2 \equiv \tau$ and $K_{12}=K_{21}\equiv K$. Since one 
is looking for a {\it symmetric} spatial configuration for these two 
$s$-wave baryons, it is the symmetric combination of the $T_i$'s that 
is required: 
\begin{equation} 
(T_1 + T_2) = -\tau K \ast (T_1 + T_2)~, 
\end{equation} 
and the requirement of vanishing Fredholm determinant at bound-state energies 
becomes equivalent to searching for zeros of the operator $(1+\tau K)$. The 
operator $\tau$ is positive definite for the attractive meson-baryon 
interactions considered in the present work, and the operator $K$ is negative 
definite at energies below threshold. Thus, if the meson-baryon interaction 
is sufficiently strong, the operator $(1+\tau K)$ will have a zero at 
a subthreshold energy. Indeed for such a fictitious $(I,J^P)=(2,2^+)$ 
$\pi NN$ system excluded by the Pauli principle, and using $\pi N$ 
interaction parameters from Table~\ref{tab:piNR}, Eq.~(\ref{eq4}), 
we get a bound state with binding energy $B_{\pi NN}=29$~MeV. 

For {\it physical} $\pi NN$ and $\pi \Lambda \Lambda$ systems, with 
symmetric spin-isospin configurations chosen, the Pauli exclusion principle 
requires that the spatial configuration be {\it antisymmetric}, leading 
to the requirement of finding zeros of the operator $(1-\tau K)$. Since 
$\tau K$, for the meson-baryon interactions considered here, is negative 
definite below threshold, this means that the operator $(1-\tau K)$ assumes 
values higher than one below threshold, which is commonly interpreted in 
terms of three-body {\it repulsion}. It is unlikely that adding secondary 
interaction channels into this schematic calculation will change the 
conclusion that no bound states are expected for $\pi BB$ systems with 
two identical ${\frac{1}{2}}^+$ baryons.

\end{document}